\documentclass[aps,prd,a4paper,superscriptaddress,nofootinbib,
showpacs,showkeys,amsfonts,amssymb,amsmath]{revtex4}
\usepackage{amssymb,latexsym}
\usepackage{amsmath, amsthm}
\usepackage{amscd}
\usepackage{amsthm}
\usepackage{times}
\usepackage{epsfig}
\usepackage{psfrag}
\usepackage{graphicx}
\usepackage{amssymb,latexsym}
\begin{document}
\title{Shell-crossings in Gravitational Collapse}
\author{Pankaj S. Joshi} \email{psj@tifr.res.in}
\affiliation{Tata Institute of Fundamental Research, 
Homi Bhabha road, Colaba, Mumbai 400005, India}
\author{Ravindra V. Saraykar} \email{ravindra.saraykar@gmail.com}
\affiliation{Department of Mathematics, R.T.M. Nagpur 
University, Nagpur, 440033, India}

\begin{abstract}
An important issue in the study of continual 
gravitational collapse of a massive matter cloud in 
general relativity is whether shell-crossing singularities 
develop as the collapse evolves. We examine this here 
to show that for any spherically symmetric collapse in general, 
till arbitrarily close to the final singularity, there is 
always a finite neighborhood of the center in which 
there are no shell-crossings taking place. It follows that 
in order to study the visibility or otherwise of the 
ultra-dense region close to the final singularity of collapse 
where physical radius of the matter cloud shrinks to an
arbitrarily small value, we can always consider without 
loss of generality a collapsing ball of finite comoving 
radius in which there are no shell-crossings. 
\end{abstract}
\maketitle
%\section{Introduction}

Black holes have found many applications in
modern astronomy and astrophysics today, especially 
for very high energy observed phenomena in the universe.
Such black holes would be possibly created when large 
massive stars collapse under the pull of their own 
gravity on exhausting their internal nuclear fuel. 
One of the key important problem in this connection is 
that of black hole formation in gravitational collapse 
of a massive star. The general theory of relativity predicts 
that the final outcome of a continual collapse is a 
spacetime singularity, where the mass-energy densities, 
spacetime curvatures and other physical quantities 
blow up to become arbitrarily large. But general relativity 
does not predict that such a singularity will be necessarily 
covered in an event horizon, forming a black hole. 

Therefore the question is, when a massive star 
collapses under its own gravity if it would necessarily 
form a black hole. What is needed to resolve
this profound issue at the heart of modern black hole 
physics and its astrophysical applications is a careful 
study of the collapse phenomena within the framework 
of gravitation theory. Such a treatment of dynamical 
collapse would be essential to determine the final fate 
of a massive collapsing star which shrinks catastrophically 
under the force of its own gravity.

The continual gravitational collapse of a 
matter cloud in the framework of general relativity 
was first investigated by Oppenheimer and Snyder, 
and Datt [1,2]. 

They considered a homogeneous spherical star with 
vanishing internal pressures and zero rotation. It was shown 
that under these idealized conditions, the cloud collapses
simultaneously to a spacetime singularity, covered 
within an event horizon that develops before the singularity
as the collapse proceeds. Thus a black hole develops in the 
spacetime which hides the singularity from any faraway 
observers. This classic picture became the foundation of 
an extensive theory and astrophysical applications of modern 
black hole physics further to the cosmic censorship 
conjecture [3], namely that all realistic 
massive stars undergoing a continual gravitational collapse 
have the same qualitative behaviour,
i.e. the spacetime singularity of collapse must be always 
covered by an event horizon of gravity and 
hidden within a black hole.

While extensive theory and applications of black 
hole physics developed in past decades based on this assumption, 
the cosmic censorship conjecture remained unproved despite 
efforts of many decades. On the other hand, many collapse 
scenarios have been found in gravitation theory in past years, 
where a dynamically evolving gravitational collapse would 
terminate in a naked singularity without horizon, rather 
than a black hole. Therefore much effort has been devoted 
towards understanding and analyzing the final 
fate of physically realistic gravitational collapse 
scenarios to determine under what situations the collapse 
ends in a black hole and when a naked singularity 
will form, not covered by an event horizon  
in violation to the censorship conjecture (see e.g.
Refs [4-14, 15] 
and references therein).
The main point in these studies is to determine the
nature of the spacetime singularity of collapse 
and the super ultra-dense regions near the same 
where the physical radius of the cloud goes to a vanishingly
small value, in terms of its visibility or otherwise 
to external observers.

Towards such a purpose the dynamical collapse 
evolution of a massive matter cloud is to be studied and 
investigated using Einstein equations.
In this connection, an important issue in the theory of 
gravitational collapse is that of formation of the so 
called `shell-crossing singularities' that could 
develop, as opposed to the final genuine 
singularity where the entire cloud collapses to a zero 
physical radius. At a shell-cross, nearby shells of matter 
intersect creating momentary density singularities 
where some of the curvature scalars could blow up 
[16-19]. 
While these are known to be weak 
singularities which are resolved through a suitable extension
of the spacetime (as opposed to big bang or strong curvature 
shell-focusing singularities), the coordinate system used 
to study the collapse evolution could break down at such 
shell-crossings, and so the conclusions on the final singularity 
become unclear in that case. Hence most of the gravitational 
collapse studies generally assume that there are no 
shell-crossings present within the evolving matter cloud.

Our purpose here is to show that in fact such an 
assumption is not needed, and that given any spherically 
symmetric collapse model, there always exists a finite 
neighborhood of the central line $r=0$ such that 
there are no shell-crossings present in this 
ball of coordinate radius $[0, r_1],$ through out 
the evolution of collapse upto any arbitrarily close 
epoch to the final curvature singularity. Therefore, 
given any physically realistic collapsing star with a 
boundary at the coordinate radius $r=r_b$, if we are to 
study the visibility or otherwise of the super ultra-dense 
region very close to the central shell-focusing singularity 
at $r=0$, we can always consider a neighborhood $[0,r_1]$ 
of the central line and examine the collapse evolution 
all the way without encountering any shell-crosses. Thus 
the essential nature of the final singularity can be 
determined without bothering for any shell-crosses 
in between.

We note that shell-crosses have been regarded 
generally as weak singularities (see e.g. Ref. 20
and the discussion therein). 
This is because, non-spacelike geodesics falling into 
a shell-crossing would not generally get focused into a 
surface or a line. Thus the volume elements along the 
geodesics are not crushed to a zero size, unlike at 
the big bang or a strong curvature shell-focusing singularity 
at the center of the Schwarzschild spacetime. 
In that sense, the material objects hitting a 
shell-crossing are not crushed out of existence, which 
is the signature of a genuine spacetime singularity. 
Also, in real astrophysical objects when densities are
high then pressure gradients are present which are important
and these may prevent the occurrence of shell-crossings. 
Actually, the shell-crossings seen sometimes in the 
Lemaitre-Tolman-Bondi models are not general enough, 
and it is believed that these could be a zero-pressure 
limit of an acoustic wave of high but finite density.

The shell-crossing singularities were studied also by
Szekeres and Lun [21], 
who considered Newtonian and general 
relativistic spherically symmetric dust solutions and suggested the 
following criterion for a singularity to be classified 
as a shell-cross: (1) All Jacobi fields have finite limits 
(in an orthonormal parallel propagated frame) as they 
approach the singularity. (2) The boundary region forms 
an essential $C^2$ singularity which is $C^1$ regular, 
that is, it can be transformed away by a $C^1$ coordinate 
transformation. This allows one to think that such a 
shell-cross can be possibly avoided if the 
shapes of the arbitrary functions available in the 
geometry are properly chosen. Generally there are two ways 
suggested to avoid the shell-crossings. The first one 
is by setting up the functions so that the spatial 
derivative of the physical radius $R'$ does not vanish 
throughout the collapsing cloud, and the second is by 
setting up the functions so that $R'= 0$ only at 
those locations $r = r_s$ where $M' = 0$, and 
$Lim |M'/R'|$ 
is finite, as $r$ approaches $r_s$ [22].

We show below that the first situation above 
is in fact always realized in a general spherically 
symmetric collapse, in a finite neighborhood of 
the central line, upto any epoch arbitrarily close
to the final central singularity of collapse.

The general spherically symmetric line element
describing the collapsing matter cloud can be written
as,
\begin{equation}\label{metric}
ds^2=-e^{2\nu(t, r)}dt^2+e^{2\psi(t, r)}dr^2+R(t, r)^2d\Omega^2,
\end{equation}
with the stress-energy tensor for a generic matter 
source given by,
\begin{equation}
T_t^t=-\rho; \; T_r^r=p_r; \; T_\theta^\theta=T_\phi^\phi=p_\theta.
\end{equation}
The above is a general scenario, in that it involves 
no assumptions on the form of the matter or the equation 
of state.
The dynamical evolution of the collapsing cloud 
and its final endstate is governed by the Einstein equations, 
which we study to understand the nature of the 
singularity of collapse.
The visibility or otherwise of the singularity 
or the region close to it is 
determined by the behavior of the apparent horizon 
in the spacetime, which is the boundary of the trapped 
surface region that develops as the collapse progresses.

We define a scaling function $v(r,t)$ by 
the relation,
\begin{equation}\label{v}
R=rv.
\end{equation}
where $R$ is the physical radius of the cloud.
The Einstein equations are then written as,
\begin{eqnarray}\label{p2}
p_r&=&-\frac{\dot{F}}{R^2\dot{R}} \; ,\\ \label{rho2}
\rho&=&\frac{F'}{R^2R'} \; ,\\ \label{nu2}
\nu'&=&2\frac{p_\theta-p_r}{\rho+p_r}\frac{R'}{R}-
\frac{p_r'}{\rho+p_r} \; ,\\ \label{G2}
2\dot{R}'&=&R'\frac{\dot{G}}{G}+\dot{R}\frac{H'}{H} \; ,\\
\label{F2}
F&=&R(1-G+H) \; ,
\end{eqnarray}
where the functions $H$ and $G$ are defined as,
\begin{equation} \label{HG}
H =e^{-2\nu(r, v)}\dot{R}^2 , \; G=e^{-2\psi(r, v)}R'^2.
\end{equation}
In the above, we have five equations in
seven unknowns, namely $\rho,\; p_r, \; p_{\theta}, \; 
R,\; F,\; G,\; H$. Here $\rho$ is the mass-energy density, 
$p_r$ and $p_\theta$ are the radial and tangential pressures, 
$R$ is the physical radius for the matter cloud, and 
$F$ is the Misner-Sharp mass function. With the definitions 
\eqref{v} and \eqref{HG},  we can substitute
the unknowns $R, H$ with $v, \nu$. Without loss of generality, 
the scaling function $v$ can be set $v(t_i, r)=1$ at the 
initial time $t_i=0$ when the collapse commences.
It then goes to zero at the final spacetime singularity $t_s$, 
which corresponds to $R=0$, {\it i.e.} $v(t_s, r)=0$.
This amounts to the scaling $R=r$ at the initial epoch,
which is an allowed freedom. The collapse condition 
here is the requirement that $\dot R<0$ throughout 
the evolution, which is equivalent to $\dot{v}<0$.

With this formalism for spherical collapse,
we can consider now a continual collapse, with $v$ 
evolving from $v=1$ (initial epoch) to $v=\epsilon$, 
the later being an arbitrarily small positive quantity 
corresponding to the region of ultra-high density 
and pressure arbitrarily close to the final singularity 
epoch $v=0$.

Then what we show below is: For any arbitrarily late
stage of collapse, corresponding to $v=\epsilon$, where 
$\epsilon>0$ is an arbitrarily small quantity, there 
always exists a $\delta$ such that in the $[0,r_1=\delta]$ 
neighborhood of the central line $r=0$, there are 
no shell-crossing singularities.

Thus the point is, given any general spherical collapse,
there is always a value $r=r_1$, such that there are no 
shell-crosses in the cloud when the cloud boundary is 
$[0,r_1]$. Then we can consider the collapse all the way 
arbitrarily close to singularity to examine the visibility 
or otherwise of the arbitrarily small collapsed ball of 
ultra high densities and pressure without problem of 
shell-crossings developing in between. Since there is no 
scale in the problem, such a $[0,\delta]$ cloud for 
the range of comoving coordinate $r$ is as good as any 
other finite cloud in principle. So we can choose the boundary 
of the compact collapsing object within this comoving 
radius $r_1$, and for such a cloud there 
are no shell-crossings occurring so we can examine the 
nature of the singularity and the region very close to 
it in terms of its visibility or otherwise.  
Thus, given any arbitrary small $\epsilon > 0$, there are 
no shell-crossing singularities occurring in a finite neighborhood 
of the central line $r=0$, that is, in a finite collapsing cloud
for the entire evolution in the range $[1, \epsilon]$ for $v$. 
So we can regularly evolve the collapse from the regular 
initial surface $v=1$ to any $v=\epsilon$, arbitrarily close 
to the final shell-focusing singularity, without any 
shell-crossing occurring in the cloud of a finite size 
around the central line.

We consider a compact collapsing matter cloud, which has 
the boundary $r=r_b$, {\it i.e.} the radial coordinate $r$ 
is in the interval $[0,r_b]$. We consider a function $J$ 
defined on the domain $D = [0,r_b] \times  [\epsilon,1]$ as 
$J(r,v) = v'$. Since the metric function $v(r,t)$ is $C^2$ 
at all regular spacetime points in the variables $r$ and $t$, 
so $J(r,v)$ is a $C^1$ function of $r$ and $v$, and 
hence a continuous function.

We note that with the scaling $R=r$ or $v=1$ at $t=0$, 
we have $R'=v+rv'$. The metric function $v(r,t)$ is 
$C^2$ or $J(r,v)$ is $C^1$ at all regular spacetime 
points. Firstly, note that at the initial surface we have 
$v=1$ and the metric functions are $C^2$. Also, at the
central line $r=0$, we have $R'= v + rv' = v$, and so $R'$ 
is always positive and finite at the center, so no shell-cross 
singularity occurs at the central line of the cloud through 
out the collapse. Also, on any $v=const.$ surface, as we move 
away from the center, there is no shell-cross if $v'=0$
or $v'>0$. If $v'<0$, even then there is no shell-cross 
on that surface till at least some finite value $r_1>0$, 
{\it i.e.} in the interval $[0,r_1]$. It follows that at the 
initial surface the metric functions are $C^2$ and continue 
to be so at the center line without any shell-crosses, and 
also in a certain neighborhood of the same as seen above, 
till the final shell focusing singularity is reached 
at $v=0$.

Since the domain $D$ is compact, $J$ is bounded and hence 
there is a positive number $M$ such that $|J(r,v)| \le  M$ 
for all $(r,v)$ in $D$. In other words, $M$ is the 
supremum of $|J(r,v)|$ taken over the domain $D$. We now 
show that whatever is the sign of $v'$, if we take 
$\delta = \frac{\epsilon}{M}$, then, for all $r$ such that 
$0<r<\delta$, the quantity $R'$ is always positive.
In the following, $\epsilon$ is to be taken small enough 
so that $\frac{\epsilon}{M} < r_b$, {\it i.e.} we remain 
within the cloud.

We consider now three cases as below:

{\it Case 1}: Firstly, we consider the case 
when $v' > 0 $. In this case, obviously, since 
$v > \epsilon > 0 $ and $r > 0 $, we have the 
minimum of $R' = v + rv' > \epsilon > 0 $, and 
so there are no shell-crossings throughout 
the collapse evolution in this case.

{\it Case 2}: Now suppose $v' < 0 $, {\it  i.e.} the 
function $J(r,v)$ is negative throughout the domain $D$. In this 
case, considering $M$ and $\delta$ as above we see that for 
$0 < r < \delta $, we get (since $|v'| = - v' $ and since 
$|v'| < M $), $r< \frac{\epsilon}{M} < -\frac{v}{v'}$.
Hence, ${-rv'} < v $ or $v+{rv'} > 0 $.
{\it i.e.} $R'$ is positive throughout, thereby avoiding 
the shell-crossings again.

{\it Case 3}: Finally, it may be the case that the 
function $J(r,v)$ takes zero or positive values for some 
points $(r,v)$ in the domain $D$, and takes negative values for 
remaining points in $D$. For those values of $r$ and $v$ 
for which $J(r,v)$ takes zero or positive values, $R'$ 
is positive as shown in {\it Case 1}.
Vanishing of $J(r,v)$ at any point means $v$ is a positive 
constant for those values of $r$, which is the case, 
for example, at the initial epoch where $v = 1$. 

In order to consider for the values $(r,v)$ where 
$J(r,v) < 0$, let $D_1$ denote the subset of $D$ which
consists of all such $(r,v)$ points. Now let $M_1$ be the 
supremum of $|J(r,v)|$ taken over the set $D_1$.
Then $M_1 \le M$, and hence $\frac{1}{M} \le \frac{1}{M_1}$.
Thus, for those $(r,v) \in D$ for which $J(r,v)<0$, and
for all $r < \delta= \frac{\epsilon}{M}$, we get,
$r < \frac{\epsilon}{M} \le < \frac{\epsilon}{M_1} 
< \frac{v}{|J(r,v)|} = \frac{v}{|v'|} = \frac{v}{-v'}$
on $D_1$. This implies, as in the {\it Case 2}, 
that for $r < \delta$, we have $R' > 0$, and therefore 
the shell-crossings are avoided.

It follows that  whatever is the behavior of 
$J(r,v)$, it is always possible to choose a neighborhood 
of the central line $r=0$ in which shell-crossings 
can be avoided. One can also consider this in the 
following alternative way. Since the metric function 
$R$ is a $C^2$ function, it follows that $R'$ is a $C^1$ 
function of $(r, t)$ or $r$ and $v$, and hence is 
continuous in both $r$ and $v$. Then we can give 
a continuity argument to ensure that we have $R'$ positive 
in the neighborhood of $r = 0$. Along $r=0$, as 
long as $v > 0$, clearly $R'$ is positive. Hence, by 
continuity of $R'$, there will be a neighborhood of 
$r=0$ where $R'$ remains positive, irrespective of 
the sign of $v'$. What we have shown above is that 
a finite neighborhood or a finite radius ball around 
the central line $r=0$ exists where there are 
no shell-crossings taking place.

We can consider, as an illustration, how the 
above works for the Lemaitre-Tolman-Bondi dust collapse 
models, which have been widely studied by 
many authors 
[23-28].

In this case, where we consider marginally bound collapse,
the Einstein field equations above can be written down 
with $p_r=p_\theta=0$ and these are solved to obtain,
\begin{equation}
R(r,t) = [ {\frac{9}{4}} F(r) (t-a(r))^2 ] ^ {1/3}, 
\end{equation}
or,
\begin{equation}
t- a(r)  = -\frac{2}{3} \frac {R^{3/2}}{\sqrt{F(r)}}
\end{equation}
where  $F(r)$ is the mass function which is always 
positive in the interval $[0,r_b]$, with $a(r)$ being
another function of integration. Clearly, $R$ is 
positive in the above interval, so $t \neq a( r )$ as 
long as $R$ and $F$ are positive. Computing the 
derivative of $R$ with respect to $r$, we get,
$R' =  [3 F' (t-a)^2]  / ( 4R^2 )  - ( 3/2 ) ( F / R^2 ) 
( t- a(r)) a' ( r )$.
Using the above expression for  
$t-a(r)$ we get,
\begin{equation}
R' = \frac{1}{3} \frac{RF'}{F}  + a'(r) \frac{\sqrt{F}}{\sqrt{R}}
\end{equation}

Since $F$ being the mass function is an increasing 
function of $r$, we have $F' > 0$  in the interval $[0,r_b]$. 
Using $(r,v)$ coordinates, where $v$ is a function of 
$(r, t)$ and where we have put  $R = r v(r,t)$, 
and $F(r) = r^3 M(r)$, we get,  
\begin{equation}
v' =  \frac{v}{3} \frac{M'}{M}  +  a' ( r ) \frac{\sqrt{M}}{\sqrt{v}}
\end{equation}
We get a supremum for $|v'|$, and we use the above 
to get a $\delta > 0$ such that for $0 \leq  r \leq \delta$, $R'$ 
is strictly positive, thereby avoiding shell-crossings. 
Towards this end, noting that $\epsilon \leq v \leq 1$ 
gives $( 1 /\sqrt{v} )  \leq  ( 1 / \sqrt{\epsilon} )$.
Since functions $M(r)$ and $a(r)$ are at least 
$C^1$, and $M$ is non-zero in the interval $[0, r_b]$, 
so the functions  $M'/ M$   and   $a' ( r ) \sqrt{M} $ 
are continuous on the compact interval $[0, r_b]$, and 
hence bounded.  Thus, let $ |M'/M|   \leq  K_1$ and   
$|a' (r)  (\sqrt{M} )|  \leq  K_2$  over the above interval.
So we get $|v'|  \leq  K$ on $[0,r_b]$,   where  
$K = (1 / 3 ) K_1  + ( K_2 ) /  \sqrt{\epsilon}$. This gives 
the desired $\delta$ where  $\delta =  \epsilon/ K$ 
and we can carry out the proof of showing that 
$R' > 0$ as before.

We thus see that for a collapsing matter cloud, 
upto any arbitrarily small value $v=\epsilon$ corresponding 
to any very late stage in collapse, there is always a 
finite radius $\delta$ around the center throughout 
the collapse such that there are no shell-crossings in 
that ball. So if we are interested in examining the visibility 
or otherwise locally of any arbitrarily small collapsing 
ball just before the final central shell-focusing singularity, 
that can be done without bothering about the shell-crosses 
all the way, which do not exist in a $\delta$ comoving 
radius around the center of the collapsing cloud.
Physically this is interesting, because as the collapse 
progresses, at a certain very late stage one would expect 
the classical relativity to break down and quantum gravity 
takes over. So if we call that stage of collapse to be 
$v=\epsilon$, a very small quantity, in reality we are  
interested in examining the visibility or otherwise of 
such an extreme late stage of collapse only, or of the 
quantum ball that formed in collapse where quantum 
gravity takes over at a very late stage.

\end{document}